\documentclass[twocolumn,superscriptaddress,aps,preprintnumbers,amsmath,amssymb,sort&compress,nofootinbib]{revtex4-2}
 
\usepackage{amsfonts,amsmath,amssymb,ascmac,bm,tensor}
\usepackage{fnpct} 
\usepackage{comment}
\usepackage{ifpdf}
\usepackage{slashed}
\usepackage{color}
\usepackage[mathscr]{eucal}
\usepackage[utf8]{inputenc}
\usepackage{physics}
\usepackage{cancel}

\usepackage{mathtools}

\ifpdf
  \usepackage{graphicx}     
  \usepackage[bookmarksopen,colorlinks=true,linkcolor=bblue,citecolor=bblue,urlcolor=ppink]{hyperref}
\else     
\fi

\definecolor{red}{rgb}{1,0,0}
\definecolor{darkred}{rgb}{0.6,0,0}
\definecolor{darkgreen}{rgb}{0.992447,0.623778,0.034597}
\definecolor{ppink}{rgb}{1,0.4,0.4}
\definecolor{bblue}{rgb}{0.284602,0.317763,0.963947}
\definecolor{purple}{rgb}{0.5 ,0, 0.7}

\newcommand{\ee}{\text{e}}
\newcommand{\Pl}{\text{Pl}}
\newcommand{\GW}{\text{GW}}

\makeatletter
\newcommand\footnoteref[1]{\protected@xdef\@thefnmark{\ref{#1}}\@footnotemark}
\makeatother

\allowdisplaybreaks[1]

\begin{document}


\title{
Bound on Induced Gravitational Waves during Inflation Era
}

\author{Keisuke Inomata}
\affiliation{Kavli Institute for Cosmological Physics, The University of Chicago, Chicago, IL 60637, USA}

\begin{abstract}
\noindent
We put the upper bound on the gravitational waves (GWs) induced by the scalar-field fluctuations during the inflation.
In particular, we focus on the case where the scalar fluctuations get amplified within some subhorizon scales by some mechanism during the inflation.
Since the energy conservation law leads to the upper bound on the energy density of the scalar fluctuations, the amplitudes of the scalar fluctuations are constrained and therefore the induced GWs are also.
Taking into account this, we derive the upper bound on the induced GWs.
As a result, we find that the GW power spectrum must be $\mathcal P_h \lesssim \mathcal O(\epsilon^2 (k/k_*)^2)$ up to the logarithmic factor, where $\epsilon$ is the slow-roll parameter and $k_*$ is the peak scale of the scalar-field fluctuations.
\end{abstract}

\date{\today}
\maketitle

\section{Introduction}

The cosmological perturbations are one of the most powerful tools to investigate the early universe.
In the inflationary scenario, the perturbations originate from the quantum fluctuations of fields during the inflation era and therefore contain information about inflation models.
Depending on the behavior under the spatial coordinate transformation, the cosmological perturbations can be classified into three types: scalar perturbations, vector perturbations, and tensor perturbations.
In particular, scalar perturbations have been the most important quantities among the three for several decades with the measurement of the anisotropies of the cosmic microwave background (CMB) and the large scale structure (LSS)~\cite{Aghanim:2018eyx}, which originate from scalar perturbations.
On top of scalar perturbations, tensor perturbations have attracted a lot of attention recently because they can behave as gravitational waves (GWs), which have been directly detected by the LIGO-Virgo collaborations~\cite{Abbott:2016blz}.
Although the GWs detected so far originate from the mergers of black holes (BHs) and neutron stars, GWs associated with the cosmological perturbations could be detected by future observations.

In this paper, we focus on the GWs induced by scalar perturbations. 
Scalar perturbations are coupled to tensor perturbations at second order in perturbations and therefore can be the source of GWs~\cite{tomita1967non,Matarrese:1993zf,Matarrese:1997ay,Ananda:2006af,Baumann:2007zm,Saito:2008jc,Saito:2009jt}. 
This kind of GW has been studied by many authors especially in the context of primordial black holes (PBHs)~\cite{Inomata:2016rbd,Ando:2017veq,Espinosa:2018eve,Kohri:2018awv,Cai:2018dig,Bartolo:2018evs,Bartolo:2018rku,Unal:2018yaa,Byrnes:2018txb,Clesse:2018ogk,Cai:2019amo,Wang:2019kaf,Ben-Dayan:2019gll,Tada:2019amh,Inomata:2019zqy,Inomata:2019ivs,Yuan:2019udt,Xu:2019bdp,Cai:2019elf,Lu:2019sti, Yuan:2019wwo, Chen:2019xse, Hajkarim:2019nbx,Ozsoy:2019lyy,Domenech:2019quo,Fu:2019vqc,Ota:2020vfn,Lin:2020goi,Ballesteros:2020qam,Giovannini:2020qta,Pi:2020otn,Domenech:2020kqm,Lu:2020diy,Vaskonen:2020lbd,DeLuca:2020agl,Kohri:2020qqd,Sugiyama:2020roc,Domenech:2020ers,Abe:2020sqb,Inomata:2020xad,Atal:2020yic,Domenech:2020xin,Kawasaki:2021ycf,Yuan:2021qgz,Kawai:2021edk} (see also Ref.~\cite{Domenech:2021ztg} for a recent review of the induced GWs).
PBHs can be produced when large density perturbations enter the horizon and be the candidates of dark matter (DM) and the BHs detected by the LIGO-Virgo collaborations~\cite{1967SvA....10..602Z,Hawking:1971ei,Carr:1974nx,Carr:1975qj,Bird:2016dcv,Clesse:2016vqa,Sasaki:2016jop} (see also Refs.~\cite{Sasaki:2018dmp,Carr:2020gox,Carr:2020xqk,Green:2020jor} for recent reviews).
The scenarios of the DM PBHs and the LIGO-Virgo PBHs require a large power spectrum of the curvature perturbations on small scales and therefore the induced GWs are expected to be large enough to be detected by the current and the future pulsar timing array (PTA) experiments (EPTA~\cite{Lentati:2015qwp}, PPTA~\cite{Shannon:2015ect}, NANOGrav~\cite{Arzoumanian:2020vkk}, SKA~\cite{Moore:2014lga,Janssen:2014dka}) and the future space-based GW detectors (LISA~\cite{Sathyaprakash:2009xs,Moore:2014lga,Audley:2017drz}, DECIGO~\cite{Seto:2001qf,Yagi:2011wg}, BBO~\cite{phinney2003big,Yagi:2011wg}).
Even if the power spectrum is not large enough for the PBH scenarios, the induced GWs can still be used to investigate the amplitude of the small-scale perturbations that cannot be measured through CMB and LSS due to the diffusion (or Silk) damping or the non-linear growth of the perturbations~\cite{Assadullahi:2009jc,Inomata:2018epa}.

Although the scalar perturbations induce GWs both during and after the inflation, we only focus on the GWs induced during the inflation in this paper.
In the early works, the induced GWs during the inflation have been studied in the context of small-sound-speed scalar fields because such scalar fields enhance the induced GWs~\cite{Biagetti:2013kwa,Biagetti:2014asa,Fujita:2014oba}.
Recently, it has been shown that, even in inflation models without small-sound-speed fields, the induced GWs could be large enough to contribute to the tensor-to-scalar ratio on large scales~\cite{Cai:2021yvq} and be larger than those induced by the scalar perturbations after the inflation era~\cite{Cai:2019jah,Zhou:2020kkf,Peng:2021zon,Cai:2021wzd}.
In these recent works, the inflaton or the spectator-field fluctuations get amplified by some resonance mechanisms on subhorizon scales and the enhanced fluctuations induce the large GWs during the inflation era.

In this paper, we put the upper bound on the induced GWs that are associated with the amplification of scalar-field fluctuations. 
If the amplification occurs instantaneously, the energy density of the enhanced field fluctuations cannot exceed the kinetic energy of the inflaton due to the energy conservation law~\cite{Adshead:2014sga}.
Even if the amplification occurs within a finite period, the energy density of the enhanced fluctuations cannot exceed the sum of the inflaton kinetic energy at the beginning of the amplification and the difference of the potential energies before and after the amplification.
These facts lead to the upper bound on the scalar-field fluctuations and therefore on the induced GWs during the inflation.
Since the kinetic energy of the inflaton is proportional to the slow-roll parameter $\epsilon (\equiv -\dot H/H^2)$, where $H$ is the Hubble parameter and the dot denotes the time derivative, the upper bound also depends on $\epsilon$.
In this paper, we derive the upper bound on the induced GWs with a given value of $\epsilon$.

The structure of this paper is as follows.
In Sec.~\ref{sec:basic_eq_gw}, we summarize the basic equations for the GWs induced by scalar perturbations during the inflation.
In Sec.~\ref{sec:bound_field_fluct}, we derive the upper bound on the amplitude of the scalar-field fluctuations from the energy conservation law.
Then, we derive the upper bound on the induced GWs in Sec.~\ref{sec:gw_bounds}, which is the main part of this paper.
We devote Sec.~\ref{sec:concl} to the conclusion.

\section{Basic equations for induced GWs}
\label{sec:basic_eq_gw}

In this section, we summarize the basic equations for the GWs induced by scalar perturbations during the inflation.
In the conformal Newtonian gauge, the scalar and tensor metric perturbations are given by 
\begin{align}
  \label{eq:metric_pertb_def}
  \dd s^2 = a^2 \left\{ -(1+2\Phi) \dd \eta^2 + \left[ (1- 2\Psi) \delta_{ij} + \frac{1}{2}h_{ij} \right] \dd x^i \dd x^j \right\},
\end{align}
where $a$ is the scale factor, $\Phi$ and $\Psi$ are the scalar perturbations, and $h_{ij}$ is the tensor perturbations. 
During the inflation era, there is no anisotropic stress at linear order~\cite{Weinberg:2008zzc}, which enables us to take $\Phi = \Psi$ in the calculation of the induced GWs.

To present a general discussion, we begin with the multi-field action:
\begin{align}
	S = \int \dd^4 x \sqrt{-g} \left[ -\frac{1}{2} \sum_J \partial^\mu \phi_J \partial_\mu \phi_J - V(\phi)  \right],
\end{align}
where $J$ is the index of the field.
Note that we assume that the sound speeds of the scalar fields are $c_s=1$ throughout this paper.
Then, the energy-momentum tensor can be expressed as
\begin{align}
	T_{\mu\nu} = g_{\mu\nu} \left[ -\frac{1}{2} \sum_J \partial^\lambda \phi_J \partial_\lambda \phi_J - V(\phi) \right] + \sum_J \partial_\mu \phi_J \partial_\nu \phi_J.
	\label{eq:e_m_tensor}
\end{align}
From the traceless-transverse component of the Einstein equation, we can obtain the equation of motion for the tensor perturbations as~\cite{Baumann:2007zm}
\begin{align}
	{h_{ij}}'' + 2 \mathcal H {h_{ij}}' - \nabla^2 h_{ij} = 4 \hat{\mathcal T}_{ij}^{\ \ lm} S_{lm},
	\label{eq:h_ij_eom}
\end{align}
where the prime denotes the derivative with respect to the conformal time $\eta$, $\mathcal H (\equiv a'/a)$ is the conformal Hubble parameter, and $\hat{\mathcal T}_{ij}^{\ \ lm}$ is the projection operator onto the traceless-transverse space.
Specifically, we define the conformal time as $\eta \equiv -\int^{t_\text{i,end}}_t \dd \bar t/a(\bar t)$, where $t_\text{i,end}$ is the physical time at the end of the inflation.
The source term $S_{lm}$ is given by 
\begin{align}
	S_{lm} = -2 \partial_l \Phi \partial_m \Phi - 4 \Phi \partial_l \partial_m \Phi + \frac{1}{M_\Pl^2} \sum_J \partial_l \delta \phi_J \partial_m \delta \phi_J,
\end{align}
where we have separated the scalar fields into their backgrounds and perturbations as $\phi_J(\eta,\bm x) = \bar \phi_J(\eta) + \delta \phi_J(\eta, \bm x)$.
Meanwhile, from the time-space component of the Einstein equation, we obtain~\cite{Weinberg:2008zzc}
\begin{align}
	&\sum_J \frac{{\bar \phi_J}'}{\mathcal H} \delta \phi_J = 2 M_\Pl^2 \left( \Phi + \frac{\Phi'}{\mathcal H} \right) \nonumber \\
	\Rightarrow \quad & 
	\sqrt{2 \epsilon} \, \sum_J \frac{{\bar \phi_J}'}{\sqrt{\sum_K {({\bar \phi_K}'})^2}} \delta \phi_J = 2 M_\Pl \left( \Phi + \frac{\Phi'}{\mathcal H} \right),	
	\label{eq:psi_delta_phi}
\end{align}
where we have used the relation $\sqrt{\sum_K {({\bar \phi_K}'})^2}/\mathcal H = \sqrt{2\epsilon}M_\Pl$.
Because of this relation, we can safely approximate the source term during the inflation era ($\epsilon \ll 1$) as 
\begin{align}
	S_{lm} \simeq \frac{1}{M_\Pl^2} \sum_J \partial_l \delta \phi_J \partial_m \delta \phi_J.
\end{align}

We expand the tensor perturbation with two polarization modes ($+, \times$) as 
\begin{align}
	h_{ij}(\bm x) = \sum_{\lambda = +,\times} \int \frac{\dd^3 k}{(2\pi)^3} \ee^{\lambda}_{ij}(\hat k) h^{\lambda}_{\bm k} e^{i \bm k \cdot \bm x}, 
\end{align}
where $\hat k \equiv \bm k/\abs{\bm k}$ and $\ee^{\lambda}_{ij}$ are the two polarization tensors, which satisfy
\begin{align}
	\hat k^{i} \ee^{\lambda}_{ij}(\hat k) = 0, \ \ee^{\lambda \, i}_{\ \ \  i} (\hat k) = 0, \ \ee^{\lambda \, ij}(\hat k) \ee^{\lambda'}_{\ \ ij}(\hat k) = \delta^{\lambda \lambda'}.
\end{align}
Then, we can rewrite Eq.~(\ref{eq:h_ij_eom}) as 
\begin{align}
  	{h_{\bm k}^{\lambda}}'' + 2 \mathcal H {h_{\bm k}^\lambda}' + k^2 h_{\bm k}^{\lambda} = 4 \mathcal S_{\bm k}^{\lambda},
  	\label{eq:eom_tensor}
\end{align}
where
\begin{align}
	\mathcal S_{\bm k}^{\lambda} \simeq \frac{1}{M_\Pl^2} \int \frac{\dd^3 p}{(2\pi)^3} \ee^\lambda_{ij}(\hat k) p^i p^j \sum_J \delta \phi_{J,\bm p} \delta \phi_{J,\bm k - \bm p}.
\end{align}

For later convenience, we expand $\delta \phi$ with the creation and the annihilation operators ($a(\bm k), a^\dagger(\bm k)$) as 
\begin{align}
	\delta \phi_J(\bm x,\eta) &= \int \frac{\dd k^3}{(2\pi)^3} \ee^{i \bm k \cdot \bm x} \delta \phi_{J,\bm k}(\eta) \nonumber \\
	&= \int \frac{\dd k^3}{(2\pi)^3} \ee^{i \bm k \cdot \bm x} \left[ U_J(k,\eta)a_J(\bm k) +  U^{*}_J(k,\eta) a_J^{\dagger}(-\bm k) \right],
\end{align}
where the commutation relations between the creation and annihilation operators are given by
\begin{align}
		&[a_J(\bm k), a_{J'}(\bm k')] = [a_J^\dagger(\bm k), a_{J'}^\dagger(\bm k')] = 0, \\
		&[a_J(\bm k), a_{J'}^\dagger(-\bm k')] = (2\pi)^3 \delta_{JJ'} \delta(\bm k + \bm k').
\end{align}
Here, we define the power spectrum in the superhorizon limit and the transfer function as 
\begin{align}
		\mathcal P_{\delta \phi_J}(k) &\equiv \frac{k^3}{2\pi^2} |U_J(k,\eta \rightarrow 0)|^2, \nonumber \\
		T_J(k,\eta) &\equiv \frac{U_J(k,\eta)}{U_J(k, \eta \rightarrow 0)}.
\end{align}
Then, we obtain the following expression of the expectation value:
\begin{align}
	\expval{\delta \phi_{J,\bm k}(\eta_1) \delta \phi_{J',\bm k'}(\eta_2) } = &(2\pi)^3 \delta_{JJ'} \delta (\bm k + \bm k') \nonumber \\
	&\times T_J(k,\eta_1) T^*_J(k,\eta_2) \frac{2\pi^2}{{k}^3} \mathcal P_{\delta \phi_J}(k).
\end{align}

Next, we derive the power spectrum of the tensor perturbations, $\mathcal P_h(k,\eta)$, which is related to the expectation value as
\begin{align}
	\expval{h^\lambda_{\bm k}(\eta)  h^{\lambda'}_{\bm k'}(\eta)} = (2\pi)^3 \delta^{\lambda \lambda'} \delta(\bm k + \bm k') \frac{2\pi^2}{k^3} \mathcal P_h(k,\eta).
\end{align}
Solving Eq.~(\ref{eq:eom_tensor}), we obtain 
\begin{align}
	h^\lambda_{\bm k}(\eta) = 4\int^\eta_{-\infty} \dd \eta' g_k(\eta; \eta') \mathcal S^\lambda_{\bm k} (\eta'), 
\end{align}
where $g_k$ is the Green function that satisfies 
\begin{align}
	g_k'' + 2\mathcal H g_k' + k^2 g_k = \delta(\eta - \eta').
\end{align}
Note that the prime denotes the derivative with respect to $\eta$, not $\eta'$.
Substituting $\mathcal H \simeq -1/\eta$ and imposing the causality, we obtain the concrete expression of $g_k$:
\begin{align}
	g_k(\eta;\eta') = \Theta(\eta - \eta') \frac{1}{k^3{\eta'}^2} &\left\{ k(\eta' - \eta) \cos[k(\eta' - \eta)] \right.\nonumber \\
	& \left. \quad 
	- (1+ k^2 \eta \eta') \sin[k(\eta' - \eta)] \right\}.
\end{align}
Then, we can rewrite the expectation value as 
\begin{align}
\label{eq:tensor_ex_value}
\expval{h^\lambda_{\bm k}(\eta)  h^{\lambda'}_{\bm k'}(\eta)} = 
16 \int^\eta_{-\infty} \dd \eta_1 \int^\eta_{-\infty} &\dd \eta_2 \, g_{k}(\eta; \eta_1) g_{k'}(\eta; \eta_2) \nonumber \\
 &\times \expval{\mathcal S^\lambda_{\bm k} (\eta_1) \mathcal S^{\lambda'}_{\bm k'}(\eta_2)}.
\end{align}
Assuming that $\delta \phi$ follows the Gaussian distribution\footnote{The effects of the non-Gaussianity of scalar perturbations are discussed in Refs.~\cite{Nakama:2016gzw,Garcia-Bellido:2017aan,Cai:2018dig,Unal:2018yaa,Adshead:2021hnm}.}, we express the expectation value of the source terms as 
\begin{widetext}
\begin{align}
\expval{\mathcal S^{\lambda}_{\bm k}(\eta_1) \mathcal S^{\lambda'}_{\bm k'} (\eta_2)} &= \frac{1}{M_\Pl^4} \int \frac{\dd^3 p \,\dd^3 p'}{(2\pi)^6} \ee^{\lambda \, ij}(\hat k) p_i p_j \ee^{\lambda' \, lm} (\hat k') p_l' p_m' \sum_J \sum_{J'} \expval{\delta \phi_{J,{\bm p}}(\eta_1)  \delta \phi_{J,\bm k- {\bm p}} (\eta_1) \delta \phi_{J',{\bm p}'}(\eta_2) \delta \phi_{J',\bm k'- {\bm p}'} (\eta_2)} \nonumber \\
&= (2\pi)^3 \delta(\bm k+ \bm k') \delta^{\lambda \lambda'} \frac{2\pi^2}{k^3} \frac{1}{4M_\Pl^4} \int^\infty_0 \dd p \int^1_{-1} \dd\mu \frac{k^3 p^3}{|\bm k- {\bm p}|^3} (1-\mu^2)^2 \nonumber \\
& \qquad \qquad \qquad \qquad \qquad\times  \sum_J f_J(|\bm k- {\bm p}|,p, \eta_1)  f_J^*(|\bm k- {\bm p}|,p, \eta_2) \mathcal P_{\delta \phi_J}(p) \mathcal P_{\delta \phi_J} ( |\bm k- {\bm p}| ),
\label{eq:gw_source_ss_rr}
\end{align}
where $\mu$ is defined as $\mu \equiv \bm k \cdot {\bm p}/(k p)$, and $f_J$ is given by 
\begin{align}
	f_J(k_1,k_2, \eta) = T_J(k_1,\eta) T_J(k_2,\eta).
\end{align}
We can rewrite Eq.~(\ref{eq:gw_source_ss_rr}) with $u \equiv |\bm k - {\bm p}|/k$ and $v \equiv p/k$ instead of $p$ and $\mu$ as
\begin{align}
\label{eq:gw_source_ss_rr2}
\expval{\mathcal S^{\lambda}_{\bm k}(\eta_1) \mathcal S^{\lambda'}_{\bm k'} (\eta_2)} 
&= (2\pi)^3 \delta(\bm k+ \bm k') \delta^{\lambda \lambda'} \frac{2\pi^2}{k^3}  \nonumber \\
&\ \ \times \frac{k^4}{4M_\Pl^4} 
\int^\infty_0 \dd v \int^{|1+v|}_{|1-v|} \dd u \left[ \frac{4v^2 - (1 + v^2 - u^2 )^2}{4uv} \right]^2  \sum_J f_J(u k, v k, \eta_1)  f_J^*( u k, v k, \eta_2)  \mathcal P_{\delta \phi_J}(u k) \mathcal P_{\delta \phi_J} ( v k ),
\end{align}
where we have used the relation $\mu = (1+v^2 - u^2)/2v$.
From Eqs.~(\ref{eq:tensor_ex_value}) and (\ref{eq:gw_source_ss_rr2}), we finally obtain 
\begin{align}
\label{eq:ph_express_v_u}
\mathcal P_h(k,\eta) = \frac{4}{M_\Pl^4}&\sum_J  
\int^\infty_0 \dd v \int^{|1+v|}_{|1-v|} \dd u \left[ \frac{4v^2 - (1 + v^2 - u^2 )^2}{4uv} \right]^2 |I_J(u, v, k, \eta)|^2 \mathcal P_{\delta \phi_J}(u k) \mathcal P_{\delta \phi_J} ( v k ),
\end{align}
\end{widetext}
where $I_J(u,v,k,\eta)$ is defined as
\begin{align}
  \label{eq:i_vux_def}
  I_J(u,v,k,\eta) \equiv k^2 \int^\eta_{-\infty} \dd \bar \eta \, g_k(\eta; \bar \eta) f_J(uk,vk,\bar \eta).
\end{align}
The relation between the energy density parameter and the power spectrum of the GWs is given in Appendix~\ref{app:gw_energy}.

\section{Bound on scalar-field fluctuations}
\label{sec:bound_field_fluct}

In this section, we derive the upper bound on the fluctuations of the scalar fields by taking into account the energy conservation law.
From the energy-momentum tensor (Eq.~(\ref{eq:e_m_tensor})), the energy density is given by 
\begin{align}
	\rho &= -T^0_{\ \ 0} \nonumber \\
	&=  \sum_J \left( \frac{1}{2}\partial^\mu \phi_J \partial_\mu \phi_J + V(\phi) - \partial^0 \phi_J \partial_0 \phi_J \right).
	\label{eq:rho_def}
\end{align}
From this, the energy density of the field fluctuations is given by~\cite{Adshead:2014sga} 
\begin{align}
	\rho_f &\simeq  \frac{1}{2a^2} \sum_J \expval{ (\delta \phi_J')^2 + (\partial_i \delta \phi_J)^2},
	\label{eq:rho_p}
\end{align}
where we have assumed that $\delta \phi_J$ is much larger than the vacuum (or zero-point) fluctuations.\footnote{If there is no amplification of the scalar-field fluctuations, the contribution from their vacuum fluctuations gives the correction of $\mathcal O(H^4/M_\Pl^4)$ to the first-order GW spectrum, $\mathcal P_h = 4H^2/(\pi M_\Pl)^2$~\cite{Adshead:2009cb}.}
We have also neglected the contribution from the second derivative of the potential, $(\partial^2 V/\partial \phi^2) \delta \phi^2$, for simplicity.
Strictly speaking, this second derivative term could change the value of $\rho_f$.
In particular, in the case of $-\partial^2 V/\partial \phi^2 \gg H^2$, $\rho_f$ could be much smaller than Eq.~(\ref{eq:rho_p}).
Because of this, our upper bounds could in principle be aggressive if the perturbations are large \emph{only} during the period of $-\partial^2 V/\partial \phi^2 \gg H^2$ and are not during other periods.
However, at least in the previous works about the induced GWs~\cite{Cai:2019jah,Zhou:2020kkf,Cai:2021yvq,Peng:2021zon,Cai:2021wzd}, this particular situation is not realized and therefore we do not care about this possibility in this paper. 

Here, similarly to the previous works~\cite{Cai:2019jah,Zhou:2020kkf,Cai:2021yvq,Peng:2021zon,Cai:2021wzd}, we consider the case where the fluctuation amplification occurs on some peak scale $k_*$ before its horizon exit and the peak-scale fluctuations freeze on superhorizon scales.
In this case, the energy density before the horizon exit of the peak scale can be approximated by
\begin{align}
	\rho_f	&\simeq \sum_J  \int \frac{\dd k}{k} \left(\frac{k}{a}\right)^2 |T_J(k,\eta)|^2 \mathcal P_{\delta \phi_J}(k) \quad (|\eta| \gg 1/k_*),
	\label{eq:rho_p_sub}
\end{align}
where we have approximated $|\delta \phi_{\bm k}'|^2 \simeq k^2 |\delta \phi_{\bm k}|^2$, which is valid when the amplitude changes adiabatically compared to the oscillation timescale of $1/k$. 
If the perturbation amplitude changes more rapidly than the oscillation timescale, the time derivative of the fluctuations become $|\delta \phi_{\bm k}'|^2 > k^2 |\delta \phi_{\bm k}|^2$, which leads to a stronger upper bound on the scalar-field fluctuations.
In this sense, our analysis with the assumption of $|\delta \phi_{\bm k}'|^2 \simeq k^2 |\delta \phi_{\bm k}|^2$ gives a conservative bound on the induced GWs.

The energy conservation law puts the upper bound on $\rho_f$:
\begin{align}
	\rho_f(\eta) < \rho_\text{kin}(\eta_s) + (\rho_\text{pot}(\eta_s) - \rho_\text{pot}(\eta)),
	\label{eq:energy_cons_bound}
\end{align}
where $\eta_s$ is the starting time of the scalar-field amplification and $\rho_\text{kin}$ and $\rho_\text{pot}$ are the density of the kinetic and the potential energy of the inflaton, respectively.\footnote{Throughout out this paper, we regard the inflaton direction as the direction of the background fields' evolution.
Then, the inflaton kinetic energy is given by $ \sum_J (\bar\phi'_J)^2/(2a^2)$.}
Note that we have neglected the energy loss due to the Hubble friction for simplicity, which decreases the right-hand side and makes the bound stronger.
The detail derivation of this bound is given in Appendix~\ref{app:energy_cons_bound}.
Hereafter, we assume that the potential energy does not significantly change during the fluctuation amplification, satisfying $\rho_\text{pot}(\eta_s) - \rho_\text{pot}(\eta) < \mathcal O(\rho_\text{kin}(\eta_s))$, which is the case in the previous works~\cite{Cai:2019jah,Zhou:2020kkf,Cai:2021yvq,Peng:2021zon,Cai:2021wzd}.
In this case, the upper bound on the energy density becomes
\begin{align}
	&\rho_f < \mathcal O(\epsilon H^2 M_\Pl^2), 
	\label{eq:e_cons_cond}
\end{align}
where we have used $\rho_\text{kin} = \epsilon H^2 M_\Pl^2$.
Since both the total tensor power spectrum (Eq.~(\ref{eq:ph_express_v_u})) and the total energy of the enhanced fluctuations (Eq.~(\ref{eq:rho_p_sub})) are given by the sum of each field contribution, we can derive the upper bound on the induced GWs assuming the single-field case without loss of generality.
For this reason, we consider the single-field case and omit the index of $J$ in the following.

With the bound on $\rho_f$, let us derive the upper bounds on some concrete power spectra.
First, we consider a delta-function power spectrum that exhibits a peak at $k_*$:
\begin{align}
\label{eq:ps_delta_func}	
|T(k,\eta)|^2 \mathcal P_{\delta \phi}(k) = 
A^2(\eta) \delta(\ln(k/k_*)).
\end{align}
Substituting this into Eq.~(\ref{eq:rho_p_sub}), we obtain the energy density in $|\eta| \gg 1/k_*$:
\begin{align}
	\rho_f \simeq 
	A^2(\eta) \left(\frac{k_*}{a(\eta)}\right)^2.
\end{align}
From the bound on $\rho_f$, Eq.~(\ref{eq:e_cons_cond}), we can get the condition on $A(\eta)$: 
\begin{align}
	A^2(\eta) \lesssim \epsilon \left(k_*\eta \right)^{-2} M_\Pl^2 \quad (|\eta| \gg 1/k_*).
	\label{eq:a_eta_u_b}
\end{align}
As another example, we consider the top-hat power spectrum, which is parameterized by 
\begin{align}
	|T(k,\eta)|^2 \mathcal P_{\delta \phi}(k) = B^2(\eta) \Theta(k - k_c) \Theta(k_* - k),
\end{align}
where $k_c < k_*$.
Substituting this, we obtain the energy density in $|\eta| \gg 1/k_c$:
\begin{align}
	\rho_f \simeq B^2(\eta) \left( \frac{k_*^2 - k_c^2}{2a^2}\right).
\end{align}
Then, Eq.~(\ref{eq:e_cons_cond}) leads to the condition on $B(\eta)$:
\begin{align}
	 B^2(\eta)  \lesssim 2 \epsilon \left((k_*^2 - k_c^2)\eta^2 \right)^{-1} M_\Pl^2 \quad (|\eta| \gg 1/k_c).
	 \label{eq:b_eta_u_b}
\end{align}
From the above results, we can see that, when the peak-scale perturbations are much smaller than the horizon scales, the stronger upper bound is put on the amplitude of the field fluctuations. 
This is because the energy density of the fluctuations depends on the time and the spatial derivatives of the fluctuations (see Eq.~(\ref{eq:rho_p})).

\section{Bound on induced GWs}
\label{sec:gw_bounds}

In this section, we derive the upper bound on the induced GWs using that on the scalar-field fluctuations, discussed in the previous section.
Before going to the numerical calculation, we briefly estimate the order of the upper bound on the induced GWs.
Since the upper bound on the scalar fluctuations is weakest around the horizon scales as shown in the previous section, let us estimate the upper bound on the GWs induced by the horizon-scale scalar fluctuations.
For such fluctuations, all the time and the spatial derivatives in Eq.~(\ref{eq:h_ij_eom}) become the factor of $\mathcal O(1/\eta)$, which leads to the approximate relation given by $h_{ij} \sim \delta \phi^2/M_\Pl^2$.
Since the upper bound on the horizon-scale fluctuations is $\mathcal P_{\delta \phi} \lesssim \epsilon M_\Pl^2$, the approximate relation for $h_{ij}$ gives the upper bound on the induced GWs, $\mathcal P_h \lesssim \mathcal O(\epsilon^2)$.
Note that this rough estimate has also been done in Ref.~\cite{Mirbabayi:2014jqa}.
Although this rough bound is for the horizon-scale fluctuations, we will numerically see in the following that the GW spectrum cannot be larger than this upper bound in any case.

\subsection{Extreme situation}

First, we consider the extreme situation where $\rho_f \simeq \epsilon H^2 M_\Pl^2$ is always satisfied until the peak-scale perturbations exit the horizon.
In the following, similarly to the previous section, we consider the two concrete power spectra: the delta-function and the top-hat power spectrum. 

For the delta-function power spectrum, we consider the following power spectrum and the transfer function:
\begin{align}
	|T(k,\eta)|^2 \mathcal P_{\delta \phi}(k) = \epsilon M_\Pl^2 T_b^2(\eta) \delta( \log (k/k_*)),
	\label{eq:p_phi_delta}
\end{align}
where $T_b(\eta)$ is given by 
\begin{align}
	T_b(\eta) = \left( 1 + (k_* \eta)^2\right)^{-1/2}.
	\label{eq:t_b_trans}
\end{align}
Note that we normalize $T_b(\eta) \rightarrow 1$ in the superhorizon limit ($|\eta| \rightarrow 0$).
We can see that $\rho_f \simeq \epsilon H^2 M_\Pl^2$ is satisfied in $|\eta| \gg 1/k_*$ and the enhanced fluctuations freeze on superhorizon scales.
Since there is still an ambiguity in the phase of $T(k,\eta)$, we consider the following two forms of $T(k,\eta)$.
One is the form without the oscillation,
\begin{align}
	T(k,\eta) = T_b(\eta),
	\label{eq:t_no_osc}
\end{align}
and the other is with the oscillation,
\begin{align}
	T(k, \eta) = T_b(\eta) \ee^{-i k \eta},
	\label{eq:t_w_osc}
\end{align}
where we have introduced the time-dependent phase to take into account the perturbation oscillation on subhorizon scales.
Strictly speaking, for the form without the oscillation, the expression of the energy density could change from Eq.~(\ref{eq:rho_p_sub}) by a factor of $\mathcal O(1)$ because $\delta \phi_{\bm k}$ does not oscillate with the timescale $1/k$.
However, we do not care about this modification because it does not change the order of the upper bound.
In this delta-function case, the power spectrum of tensor perturbations, Eq.~(\ref{eq:ph_express_v_u}), finally becomes 
\begin{align}
\label{eq:ph_express_v_u_delta}
\mathcal P_h(k,\eta) = \epsilon^2 
& \frac{\left(4(k_*/k)^2 - 1\right)^2}{4(k_*/k)^2} |I(k_*/k, k_*/k, k, \eta)|^2 \Theta(2k_*-k).
\end{align}

On the other hand, for the top-hat power spectrum, we consider 
\begin{align}
	|T(k,\eta)|^2 \mathcal P_{\delta \phi}(k) = 2 \epsilon M_\Pl^2 \frac{k_*^2}{k_*^2 - k_c^2} T_b^2(\eta) \Theta(k-k_c) \Theta(k_* - k),
	\label{eq:p_phi_tophat}
\end{align}
where $T_b$ is given by Eq.~(\ref{eq:t_b_trans}).
We can see $\rho_f \simeq \epsilon H^2 M_\Pl^2$ in $|\eta| \gg 1/k_c$ and the perturbations freeze after the peak scale $k_*$ exits the horizon.
Even in this case, we consider the two forms of $T(k,\eta)$, given by Eqs.~(\ref{eq:t_no_osc}) and (\ref{eq:t_w_osc}).

\begin{figure}  
\centering \includegraphics[width=0.95\columnwidth]{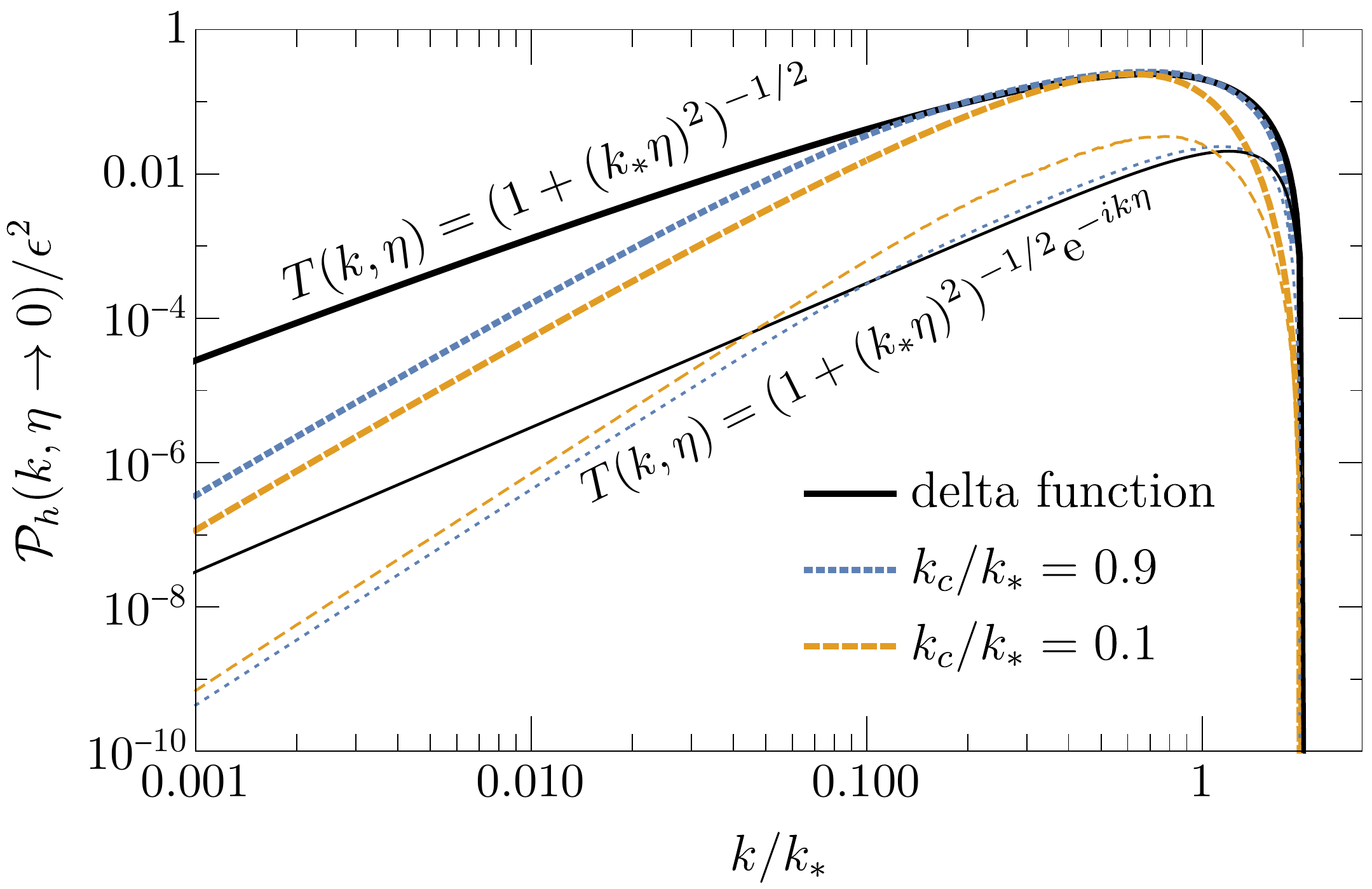}
\caption{ 
The power spectrum of the induced GWs in the extreme situation, normalized by $\epsilon^2$.
The power spectrum in the superhorizon limit ($\eta \rightarrow 0$) is shown.
The power spectrum of the scalar-field fluctuations is given by the delta function, Eq.~(\ref{eq:p_phi_delta}), for the black lines and by the top-hat function, Eq.~(\ref{eq:p_phi_tophat}), for the blue dotted and the orange dashed lines.
The thick and the thin lines show the results with the transfer functions given by Eq.~(\ref{eq:t_no_osc}) and Eq.~(\ref{eq:t_w_osc}), respectively.
}
\label{fig:gw_bounds}
\end{figure}

Figure~\ref{fig:gw_bounds} shows the power spectra of the induced GWs in the extreme case, which are calculated with Eq.~(\ref{eq:ph_express_v_u}).
In this figure, we can see that the induced GW spectrum is $\mathcal P_h \lesssim \mathcal O(\epsilon^2 (k/k_*)^2)$ up to the logarithmic factor, which we will explain below.
The peak scale of the GW spectrum is located around the smallest scale of the scalar fluctuations and the scale dependence in the large-scale limit is given by $\sim k^2$ for the delta-function case and $\sim k^3$ for the top-hat case.
The transition from $k^2$ to $k^3$ in the top-hat case is located around $k \sim (k_* - k_c)$ and this transition behavior is the same as for the spectrum of the GWs induced after the inflation~\cite{Pi:2020otn}.

Before moving on to more realistic situations, let us explain why the oscillation case with Eq.~(\ref{eq:t_w_osc}) gives the smaller GWs than the non-oscillation case with Eq.~(\ref{eq:t_no_osc}) especially in $k \ll k_*$.
The difference comes from the difference in $I$ given by Eq.~(\ref{eq:i_vux_def}). 
The integral in $k \ll k_*$ can be approximated as 
\begin{align}
	I(u,v,k,\eta) \simeq k^2 \int^\eta_{-1} \dd \bar \eta \, \left( -\frac{\bar \eta}{3} \right) T(uk,\bar \eta) T(vk,\bar \eta),
\end{align}
where the lower bound of the integral becomes $-1$ because the oscillation of the Green function suppresses the contribution from $\bar \eta < -1$.
Then, we can see that, only in the non-oscillation case, the $I$ has the logarithmic factor $\text{log}(k_*^2/k^2)$ in the late time limit ($\eta \rightarrow 0$).
Apart from that, the oscillation case has an $\mathcal O(1)$ suppression due to the oscillation of the integrand even in $\bar \eta > -1$. 
These two factors lead to the suppression of $\mathcal P_h$ in the oscillation case, Eq.~(\ref{eq:t_w_osc}), by $\mathcal O(0.1)$ around $k/k_* \sim 1$ and $\mathcal O(0.001)$ around $k/k_* \sim 0.001$, compared to the non-oscillation case. (Note again $\mathcal P_h \propto I^2$.)

\subsection{Realistic situation}

Next, we discuss a somewhat more realistic situation, where the fluctuation amplification occurs rapidly at some time and, after that, the fluctuations evolve without any sources.
To be concrete, we consider the case where the large power spectrum around $k_*$ arises at $|\eta_a| (\geq 1/k_*)$ instantaneously and satisfies $\rho_f(\eta_a) \simeq \epsilon H^2 M_\Pl^2$ and, after that, the scalar-field fluctuations evolve according to the following equation:
\begin{align}
	\delta \phi_{\bm k}'' + 2 \mathcal H \delta \phi_{\bm k}' + k^2 \delta \phi_{\bm k} = 0.
	\label{eq:delta_phi_eom}
\end{align}
Note that this equation is valid when the mass of the inflaton is negligible~\cite{Mukhanov:991646}.
Then, we can generally write down the transfer function as 
\begin{align}
		T(k,\eta,\eta_a) &\equiv \Theta(\eta - \eta_a) \left[ C (1 + ik\eta) \ee^{-ik\eta} + D (1 - ik\eta) \ee^{ik\eta} \right],
		\label{eq:general_transfer}
\end{align}
where $C$ and $D$ are constants that satisfy $C+D=1$ so that $T \rightarrow 1$ in $\eta \rightarrow 0$.
Note that this transfer function is the solution of Eq.~(\ref{eq:delta_phi_eom}) in $\eta > \eta_a$.

Similarly to the previous subsection, we consider the delta-function and the top-hat power spectrum. 
For the delta-function power spectrum, we take the following form of the spectrum in the superhorizon limit:
\begin{align}
		\mathcal P_{\delta \phi}(k) = b\, \epsilon M_\Pl^2 \left( k_* \eta_a \right)^{-4} \delta(\log(k/k_*)),
		\label{eq:p_h_delta_real}
\end{align}
where $b = (|C|^2 + |D|^2)^{-1}$.
In the case of $|k_* \eta_a| \gg 1$, we find $\rho_f(\eta_a) \simeq \epsilon H^2 M_\Pl^2$, where we have replaced $k^2|T^2|$ in Eq.~(\ref{eq:rho_p_sub}) with $(|T'|^2 + k^2 |T^2|)/2$.
Once the power spectrum is given by Eq.~(\ref{eq:p_h_delta_real}), the tensor power spectrum becomes
\begin{align}
\label{eq:ph_express_v_u_tophat}
\mathcal P_h(k,\eta) = b^2 \, \epsilon^2 
& \frac{\left(4(k_*/k)^2 - 1\right)^2}{4(k_*/k)^2} \left( k_* \eta_a \right)^{-8} \nonumber \\
&\times |I(k_*/k, k_*/k, k, \eta)|^2 \Theta(2k_*-k).
\end{align}
On the other hand, for the top-hat power spectrum, we consider the following form:
\begin{align}
	\mathcal P_{\delta \phi}(k) = b\, \epsilon M_\Pl^2 \frac{4}{( k_*^4 - k_c^4) \eta_a^4} \Theta(k - k_c) \Theta(k_* - k).
	\label{eq:p_h_tophat_real}
\end{align}
This power spectrum also satisfies $\rho_f(\eta_a) \simeq \epsilon H^2 M_\Pl^2$ in $|k_c \eta_a| \gg 1$.

\begin{widetext}

\begin{figure}
\begin{minipage}{0.49\hsize}
\begin{center}
\includegraphics[width=0.95\columnwidth]{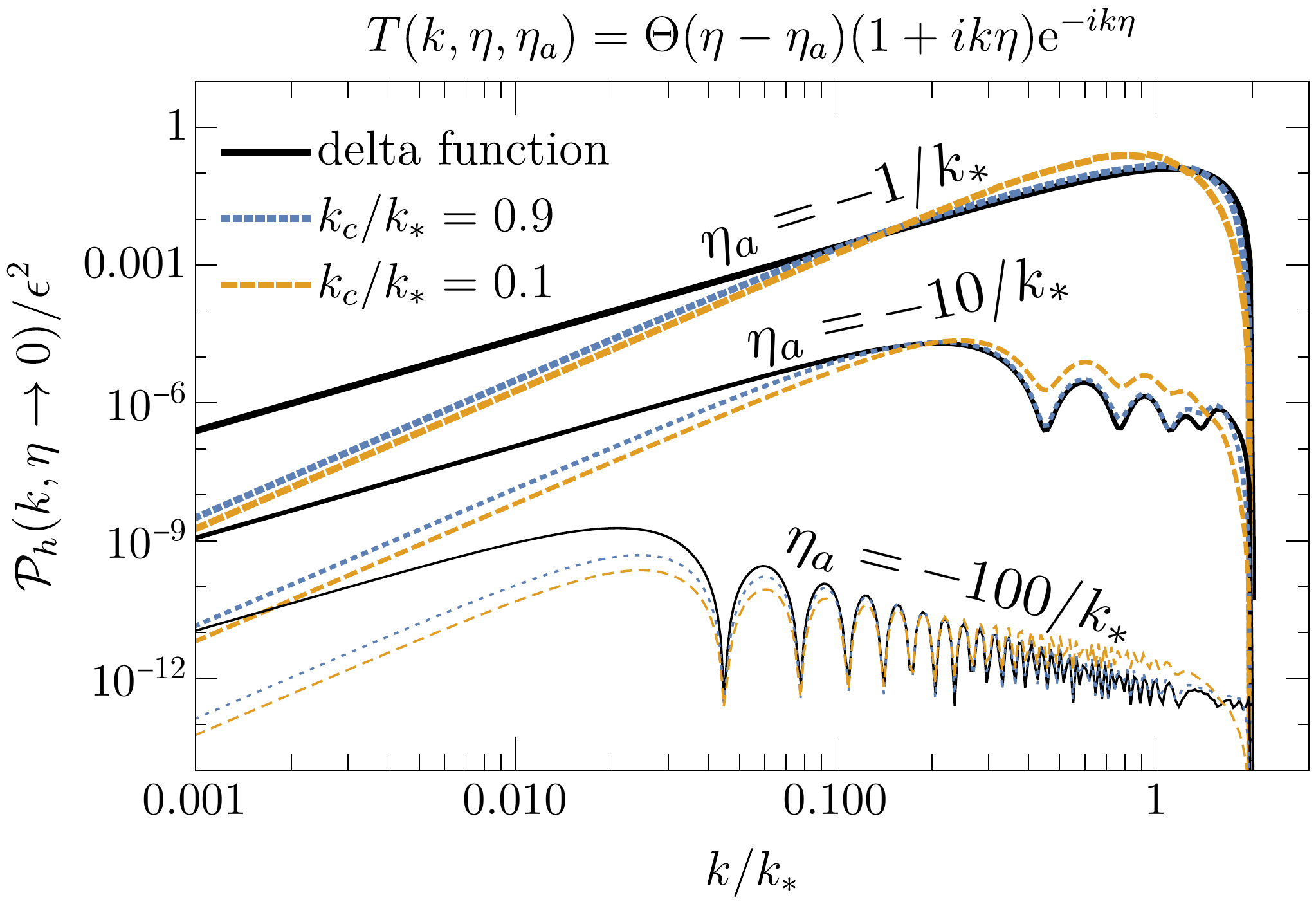}
\end{center}
\end{minipage}
\begin{minipage}{0.49\hsize}
\begin{center}
\includegraphics[width=0.95\columnwidth]{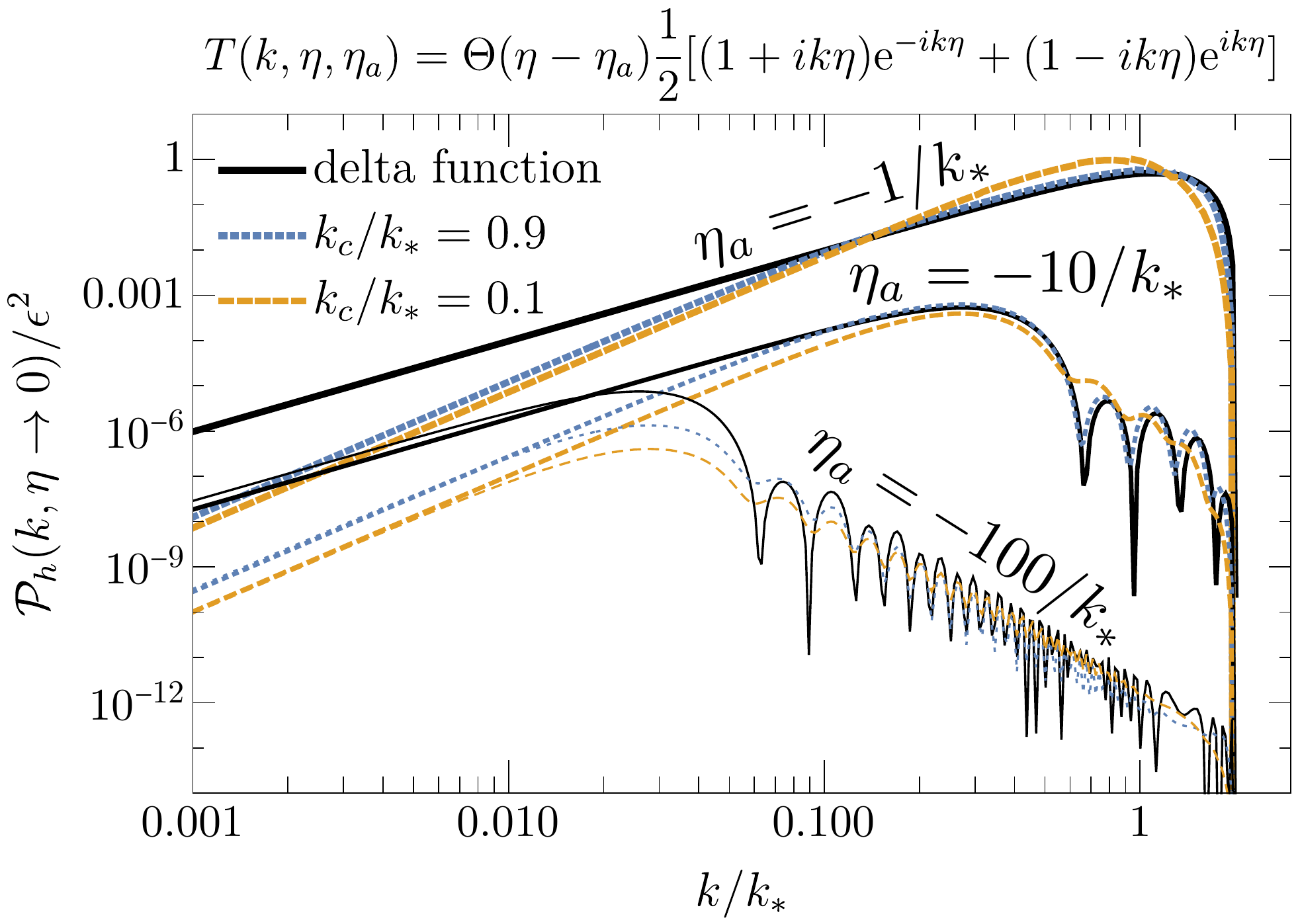}
\end{center}
\end{minipage}
\caption{
The power spectrum of the induced GWs in the realistic situation.
The transfer function is given by Eq.~(\ref{eq:general_transfer}) with $C=1, D=0$ for the left panel and with $C=D=1/2$ for the right one.
The thickness depends on the value of $\eta_a$.
The black lines are for the delta-function scalar power spectrum, given by Eq.~(\ref{eq:p_h_delta_real}), and the blue dotted and the orange dashed lines for the top-hat power spectrum, given by Eq.~(\ref{eq:p_h_tophat_real}).
}
\label{fig:gw_bounds_real}
\end{figure}
\end{widetext}

Figure~\ref{fig:gw_bounds_real} shows the numerical results in the realistic situation.
Specifically, we consider two cases: $C=1, D=0$ (left panel), and $C=D=1/2$ (right panel).
Unlike Fig.~\ref{fig:gw_bounds}, the peak of $\mathcal P_h$ is located around $k \sim \mathcal O(1/|\eta_a|)$, which corresponds to the horizon scales at the amplification of the scalar-field fluctuations.
This difference between the peak scale and the smallest scale can also be seen in Refs.~\cite{Cai:2019jah,Peng:2021zon,Cai:2021wzd}.
Apart from that, remarkably, a larger $\eta_a$ leads to a smaller GW power spectrum on the peak scale.\footnote{The connection between a smaller-scale source and a smaller upper bound on GWs can also be seen in Ref.~\cite{Giblin:2014gra}, which focuses on GWs produced after the inflation.}
This behavior originates from the fact that smaller-scale scalar fluctuations are constrained more severely, as mentioned at the end of Sec.~\ref{sec:bound_field_fluct}.
From this result, we can conclude that the induced GWs can be largest when the fluctuation amplification occurs on the scales close to the horizon and the upper bound is $\mathcal P_h \simeq \mathcal O(\epsilon^2 (k/k_*)^2)$.
We can also see that the shape of the GW spectrum depends on the transfer function.
In the case of $C=1, D=0$, the scaling of the GW spectrum on the small-scale side of the peak is $\mathcal P_h \propto k^{-2}$.
On the other hand, in $C=D=1/2$, the scaling is $\mathcal P_h \propto k^{-4}$.

\subsection{Analytical estimate of scaling relations}

In this subsection, we analytically derive the scaling relations of $\mathcal P_h$ in the realistic situation.
We can rewrite $I$, given by Eq.~(\ref{eq:i_vux_def}), as
\begin{widetext}
\begin{align}
	\label{eq:i_concrete}
	I(u,v,k,\eta) &= k^2 \int^\eta_{\eta_a} \dd \bar \eta \,  \frac{ k(\bar \eta - \eta) \cos[k(\bar \eta - \eta)] - (1+ k^2 \eta \bar \eta) \sin[k(\bar \eta - \eta)]}{k^3{\bar \eta}^2}  T(uk,\bar \eta,\eta_a) T(vk,\bar \eta,\eta_a) \nonumber \\ 
	&= \int^x_{x_a} \dd \bar x \, \frac{ (\bar x - x) \cos(\bar x - x) - (1+ x \bar x) \sin(\bar x - x)}{\bar x^2}  T(u\bar x, ux_a) T(v\bar x, v x_a),
\end{align}
\end{widetext}
where $x \equiv k \eta$, $\bar x \equiv k \bar \eta$, $x_a = k\eta_a$, and we have reparameterized the arguments of the transfer functions as 
\begin{align}
	T(ux,ux_a) &= \Theta(x-x_a) \left[ C(1 + iux) \ee^{-iux} \right. \nonumber \\
	& \qquad \qquad \qquad \qquad 
	\left. + D(1 - iux) \ee^{iux} \right]. 
\end{align}
Hereafter, we take $x \rightarrow 0$ to derive the GW spectrum in the superhorizon limit.
For convenience, we separate the contributions to $I$ depending on the coefficients as 
\begin{align}
	I(u,v,k,\eta) &= C^2 I_1 + D^2 I_1^* + 2CD I_2.
\end{align}

On the scales of $k \ll 1/|\eta_a|$, we can approximate $I_1$ and $I_2$ as 
\begin{align}
	I_1 &\simeq \int^0_{x_a} \dd \bar x \frac{uv}{3}\bar x^3  \ee^{-i (u + v) \bar x} \nonumber \\
	&\simeq \mathcal O(k^2 k_* \eta_a^3), \\
	I_2 &\simeq \int^0_{x_a} \dd \bar x \left(-\frac{uv}{3}\bar x^3 \right)  \nonumber \\
	&\simeq \mathcal O(0.1 \, k^2 k_*^2 \eta_a^4),
	\label{eq:i_delta_approx}
\end{align}
where we have used $\mathcal O(u) \simeq \mathcal O(v) \simeq \mathcal O(k_*/k) \gg 1$ and $|k_*\eta_a| \gg 1$.
Substituting this expression into Eq.~(\ref{eq:ph_express_v_u_tophat}), we obtain the order of $\mathcal P_h$ for the delta-function scalar power spectrum,
\begin{widetext}
\begin{align}
	\mathcal P_h(k(\ll 1/|\eta_a|),\eta \rightarrow 0) \sim \mathcal O \left(\frac{(C^2 + D^2)^2}{(|C|^2 + |D|^2)^2} \epsilon^2 \frac{k^2}{k_*^4 \eta_a^2} \right) + \mathcal O \left(\frac{0.01 (4 C^2D^2)}{(|C|^2 + |D|^2)^2} \, \epsilon^2 \frac{k^2}{k_*^2} \right) + \mathcal O \left( \frac{4(C^2 + D^2)CD}{(|C|^2 + |D|^2)^2} \epsilon^2 \frac{k^2}{k_*^3 \eta_a} \right),
	\label{eq:p_h_delta_approx}	
\end{align}
\end{widetext}
where the first term comes from $I_1^2$, the second one from $I_2^2$ and the third one from $I_1 I_2$.
For example, in the case of $C=1,D=0$, the first term dominates and, on the other hand, in $C=D=1/2$, the second one dominates.
Since we have ignored the coefficients, the order of this estimate is not exactly the same as the numerical results in Fig.~\ref{fig:gw_bounds_real} even in the case of $|k_* \eta_a| \gg 1$.
However, the parameter dependence of this estimate is still consistent with the numerical results.
For the top-hat power spectrum with $(k_*-k_c)/k_* \ll 1$, we can approximate the integrals of $v$ and $u$ in Eq.~(\ref{eq:ph_express_v_u}) as 
\begin{align}
	\int^\infty_0 \dd v \int^{1+v}_{|1-v|} \dd u \simeq \begin{cases}
	\mathcal O \left( \left(\cfrac{k_*- k_c}{k}\right)^2 \right) & (k > k_*-k_c)\\
	\mathcal O \left(\cfrac{k_*- k_c}{k} \right) & (k < k_*-k_c)\\
\end{cases},	
\end{align}
where we have assumed $k\ll k_*$.
Note that, in this case, all the $u$ and $v$ that appear in the integrand can be approximated as $u,v \simeq k_*/k$, similarly to the case of the delta-function power spectrum.
The point is that, while the $u$-integral becomes $\int \dd u \sim \mathcal O((k_*-k_c)/k)$ in $k > k_* - k_c$, it becomes $\int \dd u \sim \mathcal O(1)$ in $k < k_* - k_c$.
From this expression and Eq.~(\ref{eq:p_h_delta_approx}), we can see that the scale dependence of $\mathcal P_h$ changes from $k^2$ to $k^3$ at $k \sim k_* - k_c$, which is consistent with Fig.~\ref{fig:gw_bounds_real}.

On the scales of $k \gg 1/|\eta_a|$, we can approximate $I_1$ and $I_2$ as\footnote{
Strictly speaking, when $k \simeq 2 k_*$, the approximation of $I_1$ differs from Eq.~(\ref{eq:i_tophat_approx}) because $\cos\bar x\, \ee^{-i(u+v)\bar x}$ gives the term independent of $\bar x$.
This resonance effect enhances the order of $I$ by the extra factor of $k_* \eta_a$. 
However, due to the prefactor in front of $|I|^2$ in Eq.~(\ref{eq:ph_express_v_u_tophat}), this resonance effect does not enhance $\mathcal P_h$ so much.
}
\begin{align}
	I_1 &\simeq \int^0_{x_a} \dd \bar x \left( - u v \bar x\, \cos\,\bar x \right)  \ee^{-i (u + v) \bar x} \nonumber \\
	& \simeq \mathcal O(k_* \eta_a), \\
	I_2 &\simeq \int^0_{x_a} \dd \bar x  \,u v \bar x\, \cos\,\bar x \nonumber \\
	 &\simeq \mathcal O(k^{-1} k_*^2 \eta_a).
	\label{eq:i_tophat_approx}
\end{align}
Substituting this into Eq.~(\ref{eq:ph_express_v_u_tophat}), we obtain the order of the GW spectrum on the small-scale side of the peak:
\begin{widetext}
\begin{align}
	\mathcal P_h(k(\ll 1/|\eta_a|),\eta \rightarrow 0) \sim \mathcal O \left( \frac{(C^2 + D^2)^2}{(|C|^2 + |D|^2)^2} \frac{\epsilon^2 }{k^2 k_*^4 \eta_a^6} \right) + \mathcal O \left( \frac{0.01 (4C^2D^2)}{(|C|^2 + |D|^2)^2} \, \frac{\epsilon^2 }{k^4 k_*^2 \eta_a^6} \right) + \mathcal O \left( \frac{4CD (C^2 + D^2)}{(|C|^2 + |D|^2)^2} \frac{\epsilon^2 }{k^3 k_*^3 \eta_a^6} \right).
	\label{eq:p_h_delta_approx_2}	
\end{align}
\end{widetext}
Note again the first term dominates in $C=1, D=0$ and the second term dominated in $C=D = 1/2$.
This scaling relation is consistent with Fig.~\ref{fig:gw_bounds_real}.
Comparing this equation and Eq.~(\ref{eq:p_h_delta_approx}), we can see that the peak scale is $k_\text{peak} \sim \mathcal O(1/|\eta_a|)$ and the maximum peak value becomes $\mathcal P_h(k_\text{peak}) \simeq \mathcal O(\epsilon^2/(k_*\eta_a)^2)$, which is realized in the case of $C=D=1/2$. In other cases, the peak value becomes smaller than this.
This means that the peak of the GW spectrum must satisfy $\mathcal P_h(k_\text{peak}) \lesssim \mathcal O(\epsilon^2/(k_* \eta_a)^2)$.

Finally, we mention the physical interpretation of the suppression of the GW power spectrum on $k \gg 1/|\eta_a|$.
This suppression originates from two factors.
One is the redshift after the GW production. Once the GWs are produced, they behave as radiation and their amplitudes decay proportionally to the inverse of the scale factor, $h \propto 1/a$, until their horizon exits.
This effect gives the extra factor of $1/k^2$ to the $k$-dependence of the $\mathcal P_h$ on $k \gg 1/|\eta_a|$, compared to that on $k \ll 1/|\eta_a|$.
The other factor is the $k$-dependence of the GW production itself. 
In the following, we only focus on the case of $C=D=1/2$ because the situation is simpler.
From the equation of motion for GWs, Eq.~(\ref{eq:eom_tensor}), the naive relation between GWs and the source term at $\eta_a$ is given by $h_{\bm k}'' + k^2 h_{\bm k} \simeq 4 \mathcal S_{\bm k}$ on $k \gg 1/|\eta_a|$ and, on the other hand, $h_{\bm k}'' - 2h_{\bm k}'/\eta \simeq 4 \mathcal S_{\bm k}$ on $k \ll 1/|\eta_a|$.
Since the dominant contribution of $\delta \phi^2$ decays proportionally to $\eta^2$ without oscillations in the case of $C=D=1/2$, the GW-source relation gives the following GW amplitude just after their production (before the redshift effect): $h_{\bm k} \sim \mathcal O(\mathcal S_{\bm k}|_{\eta = \eta_a}/k^2)$ in $k \gg 1/|\eta_a|$\footnote{ In the case of $C=1$ and $D=0$, the oscillation of $\delta \phi^2$ changes the relation to $h_{\bm k} \sim \mathcal O(\mathcal S_{\bm k}|_{\eta = \eta_a}/(kk_*))$ in $k \gg 1/|\eta_a|$.} and, on the other hand, $h_{\bm k} \sim \mathcal O(\mathcal S_{\bm k}|_{\eta = \eta_a} \eta_a^2)$ in $k \ll 1/|\eta_a|$.
Note that the GW production occurs mainly in $\eta \sim \eta_a$ due to the $\eta^2$-decay of the source.
Then, these relations give the extra factor of $1/k^4$ to the $\mathcal P_h$ on $k \gg 1/|\eta_a|$, compared to that on $k \ll 1/|\eta_a|$.
Combining the above two factors, we finally see $\mathcal P_h \propto 1/k^4$ in $k \gg 1/|\eta_a|$ for $C=D=1/2$, which is consistent with Eq.~(\ref{eq:p_h_delta_approx_2}).

\section{Conclusion}
\label{sec:concl}

Through the interactions at second order in perturbations, GWs are induced by the scalar perturbations. 
In this paper, we have put the upper bound on the GWs induced by the scalar perturbations during the inflation era. 
To be concrete, we have focused on the case where scalar-field fluctuations get amplified by some resonance mechanism during the inflation.
The energy conservation law requires that the energy of the scalar-field fluctuations must be smaller than the sum of the inflaton kinetic energy at the beginning of the amplification and the potential energy difference before and after the amplification. 
Because of this, the amplitudes of the scalar fluctuations are upper bounded during the inflation.
Using this upper bound on the scalar fluctuations, we have derived the bound on the induced GWs.
As a result, we have found that the GWs induced during the inflation era must be $\mathcal P_h \lesssim \mathcal O(\epsilon^2 \left( k/k_*\right)^2 )$ up to the logarithmic factor, where $k_*$ is the peak scale of the fluctuation amplification.
If the induced GWs enter the horizon during the radiation-dominated (RD) era, this bound can be rewritten with the current energy density parameter as $\Omega_\text{GW} h^2 \lesssim \mathcal O(10^{-7} \epsilon^2 (k/k_*)^2)$ (see Appendix~\ref{app:gw_energy}).
This result is based on the assumption that the potential energy of the inflaton does not change significantly during the fluctuation amplification, which is the case in the slow-roll inflation with an almost constant $\epsilon$.
Even if the inflaton potential energy changes drastically during the amplification, we can still get the upper bound on the induced GWs by replacing $\epsilon H^2 M_\Pl^2$ in Eq.~(\ref{eq:e_cons_cond}) with the potential energy difference.
We should also keep in mind that this upper bound is valid only when the sound speed of the source scalar field is $c_s \simeq 1$. 
We leave the analysis in $c_s \ll 1$ for future work.

We have also found that the GW spectrum can be close to the upper bound when the scalar-field amplification occurs on scales close to the horizon.
This is because the energy density of the field fluctuations is determined by the time and the spatial derivatives of the fluctuations, which leads to a stronger upper bound on smaller-scale fluctuations.
To be concrete, we have considered the case where the amplification occurs on the subhorizon scales instantaneously at $\eta_a$, and, after that, the scalar fluctuations evolve without any sources.
Then, we have found that the GW spectrum has a peak at the scale of $k_\text{peak} \sim \mathcal O(1/|\eta_a|)$ and its peak value must be $\mathcal P_h(k_\text{peak}) \lesssim \mathcal O(\epsilon^2/(k_*\eta_a)^2)$.

Finally, we mention the implications of our results, $\mathcal P_h \lesssim \mathcal O(\epsilon^2)$. 
Note again that the following statements are valid only when $c_s \simeq 1$.

For the large-scale GWs that are the target of the CMB B-mode observations, the necessary condition for the induced GWs to be larger than the first-order primordial GWs is $\epsilon^2 > \mathcal O(\mathcal P_\mathcal R \,r_1)$, where $r_1$ is the tensor-to-scalar ratio for the first-order primordial GWs.
Note that, while $\epsilon = r_1/16$ is always satisfied in single-field inflation models, $\epsilon > r_1/16$ could be realized in multi-field inflation models~\cite{Byrnes:2006fr}.
If the large-scale GWs are observed in the future and the tensor-to-scalar ratio is determined as $r_*$, we can put the lower bound on $\epsilon$ as $\epsilon \geq \text{min} \left\{ \mathcal O(\sqrt{\mathcal P_{\mathcal R}\, r_*}), r_*/16\right\}$ even if we take into account the possibility that the observed GWs originate from the scalar-field fluctuations during the inflation~\cite{Cai:2021yvq}.

For the small-scale GWs associated with the PBH scenarios, the necessary condition for the GWs induced during the inflation to be larger than those induced after the inflation is $\epsilon > \mathcal O(\mathcal P_\mathcal R(k_\text{PBH}))$, where $k_\text{PBH}$ is the peak scale associated with the PBH mass and we have assumed that the peak scale enters the horizon during the RD era.
Note that the spectrum of the GWs induced during the RD era is given by $\mathcal P_h \sim \mathcal O(10\, \mathcal P_\mathcal R^2)$~\cite{Kohri:2018awv}.
Since the required amplitude for the PBH scenarios, such as the DM PBH and the LIGO-Virgo PBH scenarios, is $\mathcal P_\mathcal R \sim \mathcal O(0.01)$~\cite{Sasaki:2018dmp}, the necessary condition can be rewritten as $\epsilon > \mathcal O(0.01)$.
Note that this $\epsilon$ is the value around the horizon exit of the peak scale and can be different from the large-scale one that is investigated by the CMB observations.

\acknowledgments
The author thanks Peter Adshead, Misao Sasaki, and the anonymous referee for helpful comments on this paper.
The author was supported by the Kavli Institute for Cosmological Physics at the University of Chicago through an endowment from the Kavli Foundation and its founder Fred Kavli.

\appendix

\section{Energy conservation bound}
\label{app:energy_cons_bound}
In this appendix, we show the derivation of the energy conservation bound, Eq.~(\ref{eq:energy_cons_bound}).

First, we show $\expval{\rho}' \leq 0$, where $\expval{\cdots}$ represents the ensemble average.
In the following, we take the following metric: 
\begin{align}
	\dd s^2 = g_{\mu \nu} \dd x^\mu \dd x^\nu = a^2(-\dd \eta^2 + \dd x_i \dd x^i),
\label{eq:metric}	
\end{align}
where we have neglected the metric perturbations because they are suppressed by a factor of $\sqrt{\epsilon}$ compared to $\delta \phi$ during the inflation (see Eq.~(\ref{eq:psi_delta_phi})).

We begin with the energy conservation:
\begin{align}
	&T^\mu_{\ 0;\mu} = 0 \nonumber \\
	\Rightarrow \quad &
	T^\mu_{\ 0,\mu} + \Gamma^\mu_{\ \mu \lambda} T^\lambda_{\ 0} - \Gamma^\lambda_{\ 0 \mu} T^\mu_{\ \lambda} = 0,
	\label{eq:energy_consv}
\end{align}
where the Christoffel symbol is defined as 
\begin{align}
  \Gamma^\lambda_{\ \mu\nu} \equiv \frac{1}{2} g^{\lambda \rho} (g_{\rho \mu, \nu} + g_{\rho \nu, \mu} - g_{\mu \nu, \rho}).
\end{align}
In the metric given by Eq.~(\ref{eq:metric}), the Christoffel symbols become
\begin{align}
  \Gamma^0_{\ 00} = \frac{a'}{a}, \ \Gamma^i_{\ 0j} = \Gamma^0_{\ ij} = \frac{a'}{a} \delta_{ij}, \  \Gamma^i_{\ 00} = \Gamma^0_{\ 0i} = \Gamma^i_{\ jk} = 0.
  \label{eq:chris_exp}
\end{align}
From the energy-momentum tensor given by Eq.~(\ref{eq:e_m_tensor}), we can express the energy density as
\begin{align}
	\rho &= -T^0_{\ \ 0} \nonumber \\
	&=  \sum_J \left( \frac{1}{2}\partial^\mu \phi_J \partial_\mu \phi_J + V(\phi) - \partial^0 \phi_J \partial_0 \phi_J \right).
	\label{eq:rho_def}
\end{align}
Substituting these expressions into Eq.~(\ref{eq:energy_consv}), we obtain
\begin{align}
	&T^\mu_{\ 0,\mu} + 3 \mathcal H \left( T^0_{\ 0} - \frac{1}{3}T^i_{\ i} \right) = 0 \nonumber \\
	\Rightarrow \quad &
	T^\mu_{\ 0,\mu} + 3 \mathcal H \sum_J \left( \partial^0 \phi_J \partial_0 \phi_J - \frac{1}{3}\partial^i \phi_J \partial_i \phi_J  \right) = 0 \nonumber \\
	\Rightarrow \quad &
	T^\mu_{\ 0,\mu} = 3 \mathcal H \sum_J \left( (\phi_J')^2 + \frac{1}{3}\partial^i \phi_J \partial_i \phi_J  \right).
\end{align}
Since the right hand side of this expression cannot be negative, we obtain 
\begin{align}
	&T^0_{\ 0,0} + T^i_{\ 0,i} \geq 0 \nonumber \\
	\Rightarrow \quad &
	\rho' - \left(\sum_J \partial^i \phi_J \partial_0 \phi_J \right)_{,i} \leq 0.
\end{align}
Taking the ensemble average, we get 
\begin{align}
	\expval{\rho}' - \sum_J \expval{\left(\partial^i \delta \phi_J \delta \phi_J' \right)_{,i}} \leq 0.
	\label{eq:rho_with_some}
\end{align}
Since the averaged universe is homogeneous, the second term becomes zero. 
Then, we finally get 
\begin{align}
	\expval{\rho}' \leq 0.
\end{align}
This means that the energy density cannot increase in time.

Before the perturbation enhancement (when the energy density of field fluctuations is much smaller than the inflaton kinetic energy), the energy density is given by
\begin{align}
	\rho(\eta_s) &= \frac{1}{2a^2} \left[ \sum_J (\bar\phi'_J(\eta_s))^2 \right] + V(\bar \phi (\eta_s)) \nonumber \\
	&= \rho_\text{kin}(\eta_s) + V(\bar \phi(\eta_s)),
\end{align}
where note again $\rho_\text{kin}(\eta) \equiv \left[ \sum_J (\bar\phi'_J(\eta))^2 \right]/(2a^2) = \epsilon(\eta) H^2 M_\Pl^2$.
After some perturbation enhancement occurs, we must take into account the energy density of field fluctuations ($\rho_f$) as 
\begin{align}
	\rho(\eta) &= \rho_f(\eta) + \rho_\text{kin}(\eta) + V(\bar \phi (\eta)),
\end{align}
where $\eta > \eta_s$.
From $\expval{\rho}' \leq 0$, we get 
\begin{align}
	&\rho(\eta) < \rho(\eta_s) \nonumber \\
	\Rightarrow \quad &
	\rho_f(\eta) < \rho_\text{kin}(\eta_s) - \rho_\text{kin}(\eta) + V(\bar \phi(\eta_s)) - V(\bar \phi (\eta)) \nonumber \\
	\Rightarrow \quad &
	\rho_f(\eta) < \rho_\text{kin}(\eta_s) + V(\bar \phi(\eta_s)) - V(\bar \phi (\eta)),
\end{align}
where the final line is Eq.~(\ref{eq:energy_cons_bound}).

\section{Energy density parameter of GWs}
\label{app:gw_energy}

In this appendix, we derive the relation between the current energy density parameter and the power spectrum for the GWs that are induced during the inflation era and enter the horizon during an RD era.
The energy density of GWs is related to the power spectrum through~\cite{Kohri:2018awv}
\begin{align}
	\rho_\text{GW}(\eta) = \int \frac{\dd k}{k} \left( \frac{M_\Pl^2}{8} \left( \frac{k}{a} \right)^2 \overline{\mathcal P_h (k,\eta)} \right),
\end{align}
where the overline denotes the time average over the oscillations.
Here, we define the energy density per logarithmic interval in $k$ as 
\begin{align}
	\rho_\text{GW}(k,\eta) \equiv \frac{M_\Pl^2}{8} \left( \frac{k}{a} \right)^2 \overline{\mathcal P_h (k,\eta)}.
\end{align}
Then, the energy density parameter of GWs is defined by
\begin{align}
\Omega_\GW (k,\eta) \equiv \frac{\rho_\GW(k,\eta)}{\rho_{\text{tot}}} = \frac{1}{24} \left( \frac{k}{\mathcal H} \right)^2 \overline{\mathcal P_h (k,\eta)},
\label{eq:gw_cosmological_para}
\end{align}
where $\rho_\text{tot} (= 3H^2 M_\Pl^2)$ is the total energy density.
The evolution of the power spectrum during the RD era is given by
\begin{align}
	\mathcal P_h (k,\eta) = \left(\frac{\sin(k\eta)}{k\eta}\right)^2 \mathcal P_h (k,\eta \rightarrow 0),
\end{align}
where $\mathcal P_h (k,\eta \rightarrow 0)$ corresponds to the power spectrum well before the horizon entry.
Substituting this into Eq.~(\ref{eq:gw_cosmological_para}), we can see that the energy density parameter approaches the following constant value in the subhorizon limit:
\begin{align}
	\Omega_\GW(k,\eta_c) \simeq \frac{1}{48} \mathcal P_h (k,\eta \rightarrow 0),
\end{align}
where $\eta_c$ is the conformal time when $\Omega_\GW$ becomes almost constant.
Taking into account the late-time matter-dominated era, we finally obtain the current energy density parameter:~\cite{Ando:2018qdb}
\begin{align}
	\Omega_\GW (k,\eta_0) h^2 &= 0.39 \left( \frac{g_{*,c}}{106.75} \right)^{-1/3} \Omega_{r,0} h^2 \, \Omega_\GW(k,\eta_c) \nonumber \\
		&= 3.4 \times 10^{-7} \left( \frac{g_{*,c}}{106.75} \right)^{-1/3} \mathcal P_h (k,\eta \rightarrow 0),
\end{align}
where $\Omega_{r,0} h^2 (\simeq 4.18 \times 10^{-5})$ is the energy density parameter of radiation at present and $g_{*,c}$ is the effective degrees of freedom at $\eta_c$.

\small
\bibliographystyle{apsrev4-1}
\bibliography{draft_induced_gw_bound}

\end{document}